\documentclass[10pt,journal,compsoc]{IEEEtran}

\usepackage{amsbsy,amsmath,amssymb,epsfig,bbm,mathrsfs,multirow,amsthm}
\usepackage{algpseudocode}
\usepackage{subcaption,float}
\usepackage[hidelinks]{hyperref}

\usepackage[noabbrev]{cleveref}
\crefrangeformat{equation}{#3(#1)#4--#5(#2)#6}

\usepackage[linesnumbered,lined,boxed,ruled,commentsnumbered]{algorithm2e}
\usepackage{blkarray}

\usepackage[T1]{fontenc}
\usepackage[utf8]{inputenc}

\usepackage{pifont}

\usepackage{ragged2e}
\usepackage{blindtext}

\newtheorem{theorem}{\textbf{\textsc{Theorem}}}
\newtheorem{remark}{\emph{Remark}}

\DeclareMathAlphabet{\mathpzc}{OT1}{pzc}{m}{it}

\usepackage{color}

\ifCLASSOPTIONcompsoc
\usepackage[nocompress]{cite}
\else
\usepackage{cite}
\fi
\ifCLASSINFOpdf
\else
\fi

\usepackage{setspace}

\usepackage{array}
\newcolumntype{M}[1]{>{\centering\arraybackslash}m{#1}}

\usepackage{tikz}
\usetikzlibrary{decorations.pathreplacing}

\begin{document}
	
\title{Quantum Machine Learning for Secure Cooperative Multi-Layer Edge AI with Proportional Fairness}

\author{Thai T. Vu, John Le

    \IEEEcompsocitemizethanks{
        \IEEEcompsocthanksitem Thai T. Vu is with the School of Computing and Information Technology, University of Wollongong, Wollongong, Australia (email: tienvu@uow.edu.au).
        \IEEEcompsocthanksitem John Le is with the School of Computing and Information Technology, University of Wollongong, Wollongong, Australia (email: johnle@uow.edu.au).
    }
}

\IEEEtitleabstractindextext{
\begin{abstract}
This paper proposes a communication-efficient, event-triggered inference framework for cooperative edge AI systems comprising multiple user devices and edge servers. Building upon dual-threshold early-exit strategies for rare-event detection, the proposed approach extends classical single-device inference to a distributed, multi-device setting while incorporating proportional fairness constraints across users. A joint optimization framework is formulated to maximize classification utility under communication, energy, and fairness constraints. To solve the resulting problem efficiently, we exploit the monotonicity of the utility function with respect to the confidence thresholds and apply alternating optimization with Benders decomposition. Experimental results show that the proposed framework significantly enhances system-wide performance and fairness in resource allocation compared to single-device baselines.		
\end{abstract}

}

\maketitle

\IEEEdisplaynotcompsoctitleabstractindextext
\IEEEpeerreviewmaketitle

\section{Introduction}
\label{sec:introduction}


The increasing demand for real-time, intelligent services in latency-sensitive and privacy-aware applications, such as autonomous driving, industrial monitoring, and remote health diagnostics, has led to the widespread adoption of \textit{Edge Artificial Intelligence (Edge AI)}~\cite{letaief2021edge,mendez2022edge}. These systems aim to shift inference and decision-making from centralized cloud platforms to edge devices located closer to the data sources. However, resource-constrained edge nodes often struggle to run full-scale deep learning models, such as convolutional neural networks (CNNs), due to limited computational power, energy budgets, and bandwidth availability.

To address these challenges, \textit{early-exit neural networks}~\cite{liu2023resource} have been proposed to reduce the inference workload by enabling predictions to be made at intermediate layers of a model. This \textit{event-triggered inference} mechanism has been shown to significantly reduce latency and computation~\cite{teerapittayanon2016branchynet,huang2022cooperative}. Recent works have extended early-exit strategies to cooperative edge-cloud settings, where offloading decisions are jointly made based on resource conditions and prediction confidence~\cite{huang2022cooperative}. However, these methods typically assume centralized optimization and ignore fairness across users in multi-layer, multi-device environments.

Parallel to these developments, \textit{Quantum Machine Learning (QML)} has emerged as a compelling alternative for efficient learning under resource constraints. QML leverages quantum-enhanced feature representations, entanglement-based parallelism, and compact encodings to solve classification and optimization problems more efficiently than classical models in certain scenarios. Although quantum hardware is currently limited, \textit{hybrid quantum-classical models}, such as shallow \textit{quantum neural networks (QNNs)} coupled with classical layers, have proven effective on near-term devices~\cite{abbas2021power,alam2023low}. These models have shown promise in edge computing contexts, including sensor-based classification~\cite{alam2023low}, federated learning~\cite{liu2022hybrid}, and anomaly detection in IoT systems~\cite{alrubaiai2023iot}.

Despite these advances, little work has explored \textit{integrating QML into cooperative inference frameworks} that simultaneously optimize inference decisions, offloading, and wireless resource management under fairness constraints. To fill this gap, we propose a novel, communication-efficient, and quantum-aware framework for distributed inference in edge AI systems. Our contributions are summarized as follows:

\begin{itemize}
    \item We design a \textit{dual-threshold early-exit inference pipeline} that operates across user devices and edge nodes, enabling adaptive co-inference and offloading decisions based on confidence and resource states.
    \item We introduce \textit{hybrid QNNs at the user level} to perform lightweight and expressive local inference, reducing the reliance on cloud or edge servers for high-confidence samples.

    \item We formulate a \textit{joint optimization problem} for inference thresholds, user–EN assignments, and wireless resource allocation that maximizes utility while ensuring \textit{proportional fairness} among users.

    \item We develop a scalable solution based on \textit{alternating optimization} and {Benders decomposition}, leveraging monotonic properties of the threshold–utility relationship to reduce problem complexity.
    
\end{itemize}

Simulation results demonstrate that our method significantly improves system utility, fairness, and inference latency compared to existing early-exit and offloading strategies. The proposed approach presents a practical and forward-compatible path for \textit{deploying hybrid QML in multi-layer edge AI systems}, bridging quantum computing and mobile distributed intelligence.

\section{Related Work}
\label{sec:Int}

\section{System Model and Problem Formulation}
\label{sec:model_and_problem}

\subsection{System Model}
\label{sec:system_model}

\begin{figure}[h]
	\centering
	\includegraphics[scale=0.65]{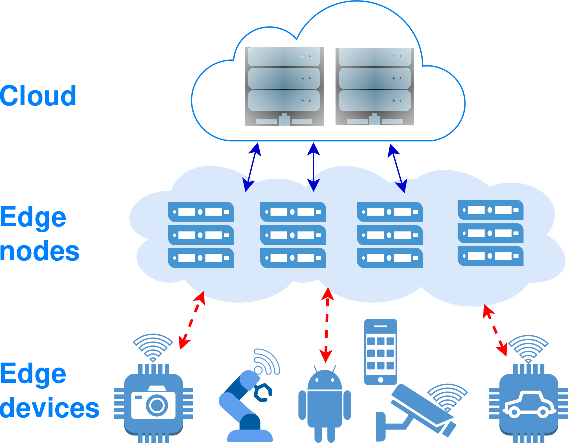}	
	\caption{Multi-layer edge computing system.}
	\label{fig:System-Model}
\end{figure}

We introduce a co-inference and training framework across a cooperative multi-layer edge computing system comprising $N$ user devices (UEs), denoted by the set $\mathbb{N}=\{1, 2, \dots, N\}$, $M$ edge nodes (ENs), represented by the set $\mathbb{M}=\{1, 2, \dots, M\}$, and a cloud server (CS), as illustrated in Fig.~(\ref{fig:System-Model}).
Each UE ($n \in \mathbb{N}$) is equipped with a light-weight convolutional neural network (CNN), which is commonly used for image recognition, to perform binary classification of events (i.e., images of incidents) into normal or critical categories. Critical events identified by UEs are offloaded to ENs ($m \in \mathbb{M}$), where each EN utilizes a more complex CNN to conduct multi-classification for detailed analysis. 

Let $\mathbb{S}=\{1,\ldots, S\}$  denote the set of possible security levels assigned to UEs, ENs, and CS, where $1$ and $S$ respectively represents the highest and lowest levels. This multi-level security model is inspired by practical frameworks, such as MapReduce~\cite{dang2019trust}, and has been adopted in several related studies~\cite{razaq2021privacy,xiao2021authentication,el2017edge}.
We assume that the event detection application at UE~$n$ requires a security level~$s_n^{\text{UE}}\in \mathbb{S}$. Accordingly, the features of critical events detected by UE~$n$ can only be offloaded to ENs that provide a security level greater than or equal to $s_n^{\text{UE}}$, thereby satisfying the application's security requirements.

\begin{figure}[h]
	\centering
	\includegraphics[scale=0.7]{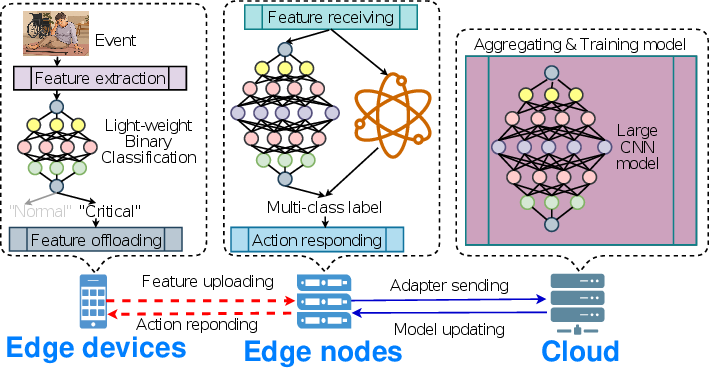}	
    \caption{Hybrid QML-based co-inference and training in a cooperative multi-layer edge system.}
	\label{fig:event_inference_and_model_updating}
\end{figure}

The CS periodically retrains the multi-classification model using updated information from ENs and UEs, subsequently updating the EN models to enhance overall system accuracy and adaptability as depicted in Fig.~(\ref{fig:event_inference_and_model_updating}). We assume that the CS and the backhaul links connecting it to the ENs are secure, thereby satisfying the security constraints of both UEs and ENs.

Assume that in a period of time, each UE~$n$ captures a sequence of independent events, denoted by $\Phi_n$. The UE~$n$ will extract the features of events as the input of the light-weight CNN for binary classification. The UE~$n$ will skip events that are detected as normal, and offload the ones determined as critical to ENs for multi-classification.
The CNN comprises multiple convolutional layers designed for feature extraction, accompanied by filters, pooling/subsampling layers, non-linear activation layers, and fully connected layers~\cite{patil2021convolutional}.

\subsubsection{Early-Exit Inference at UEs}

\begin{figure}[h]
	\centering
	\includegraphics[scale=0.62]{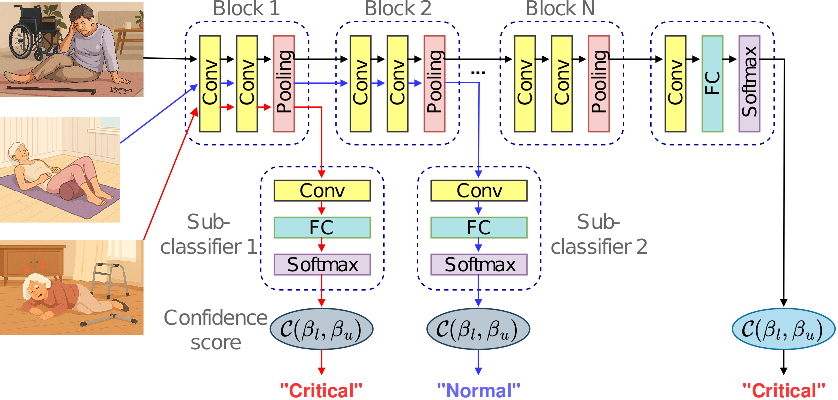}	
	\caption{Convolutional Neural Network with Early Exit.}
	\label{fig:CNN_with_early_exit}
\end{figure}

Due to different level of specific features, classification of different events faces significantly vary in complexity. For instance, events rich in features indicative of critical events can be classified as critical with minimal computational effort. Conversely, events with fewer features convey a critical one might require more extensive analysis. Hence, we employ an early exiting strategy~\cite{liu2023resource,teerapittayanon2016branchynet}, whereby at each CNN layer, if the confidence score exceeds a certain threshold, the event is immediately classified as critical. Similarly, an event can promptly be classified as normal if its confidence score falls below a defined threshold. To enhance classification performance, reduce execution time, and optimize resource consumption, we introduce a pair of confidence thresholds~\cite{gao2005bayesian,liu2023resource}, denoted by $(\alpha_n^l, \alpha_n^u)$, representing the upper and lower bounds for normal and critical events, respectively. Fig.~(\ref{fig:CNN_with_early_exit}) illustrates the CNN with early-exit strategy based on dual thresholds.

\textbf{Confidence Score.}
Let $L$ be the number of layers of the light-weight CNN at UE~$n$. 
Given the extracted feature of event~$I_k$ as input, the CNN model generates a confidence score at each layer to assess whether the event is critical or not. The confidence score $C_{q}^{(k)}$ at layer~$q$ of event~$I_k$ is defined as:
\begin{equation}
    C_{q}^{(k)} = \frac{e^{f_{q}^\text{(k),critical}}}{e^{f_{q}^\text{(k),critical}} + e^{f_{q}^\text{(k),normal}}}, \label{eq:confidence_score}
\end{equation}
where $f_{q}^{\text{(k),critical}}$ and $f_{q}^{\text{(k),normal}}$ are the outputs at layer $q$ for event $I_k$.

\textbf{Early-Exist Event Detection.}
The CNN model aims to classify each event as \textit{ critical} ($1$) or \textit{normal} ($0$) as early as possible in the network to reduce computational cost. 
At layer~$q$, if the confidence score of the event $C_{q}^{(k)} \leq \alpha_n^l$ and $\alpha_n^l < C_{t}^{(k)} < \alpha_n^u, \forall t < q$, the event is classified as normal and further processing is skipped.
Conversely, if the confidence score $C_{q}^{(k)} \geq \alpha_n^u$ and $\alpha_n^{l} < C_{t}^{(k)} < \alpha_n^{u}, \forall t < q$, the event is classified as critical, and no additional classification is needed.
If all layers up to the final one ($q \leq L$) yield confidence scores within the range $\alpha_n^l < C_q^{(k)} < \alpha_n^u$, the event is conservatively classified as normal to avoid false positives.

We define the predicted label function $\hat{y}^{(k)} \in \{$1$~(\text{critical}),~$0$~(\text{normal})\}$ of  event~$I_k$ as:

\begin{equation}
\small
\hat{y}^{(k)} =
\begin{cases}
0, &\!\!
\exists\, q \leq L, C_{q}^{(k)} \leq \alpha_n^l, \forall t < q,\ \alpha_n^l < C_t^{(k)} < \alpha_n^u \\

1, &\!\!
\exists\, q \leq L, C_q^{(k)} \geq \alpha_n^u, \forall t < q,\ \alpha_n^l < C_t^{(k)} < \alpha_n^u \\

0, &\!\!
\forall t \leq L, \alpha_n^l < C_t^{(k)} < \alpha_n^u.
\end{cases}
\label{eq:label_function}
\end{equation}

\subsubsection{Inference Energy Consumption at UEs}

During CNN inference, energy consumption is predominantly influenced by memory access rather than arithmetic operations, as accessing off-chip memory (e.g., DRAM) typically consumes over 100 times more energy than performing a single multiply–accumulate (MAC) operation~\cite{horowitz2014energy}. This observation has led to architecture-level optimizations such as ShuffleNet V2, which explicitly designs CNN blocks to minimize memory access, thus enhancing energy efficiency on user devices~\cite{ma2018shufflenet}. Consequently, it is reasonable to treat memory access operations as the main source of energy consumption and local latency.

For simplicity, we assume a uniform CNN structure across all UEs. Let $K_n$ represent the number of blocks in the CNN deployed at UE~$n$. Although an early-exit strategy is employed, we assume the energy consumption for binary classification is proportional to the total number of memory access operations. Thus, the energy consumption incurred during local inference can be expressed as follows~\cite{mao2016dynamic}: 

\begin{equation}\label{localenergy}
E_n^{l} = \gamma\sum_{i=1}^{K_n}Q^{ac}_{i},
\end{equation}
where $\gamma$ is the energy per memory access and $Q_i^{ac}$ is the number of memory access.

\subsubsection{Communication Between UEs and ENs}

In this paper, we employ Frequency Division Multiple Access (FDMA) to facilitate communication between UEs and ENs. 
Let $\boldsymbol{b} = \{b_{ij}\} \in \mathbb{R}^{N \times M}$ and $\boldsymbol{p} = \{p_{ij}\} \in \mathbb{R}^{N \times M}$ denote the allocated bandwidths and transmission powers, respectively, between UEs and ENs. 
To ensure fairness among UEs, the allocated bandwidth and transmission power are constrained by upper bounds $(b^{\text{max}}, p^{\text{max}})$~\cite{wang2016resource, chen2019energy}. 
We have the upper bound power and communication resource constraints:
\begin{equation}
    \label{eq:bandwidth_contraint}
    b_{ij} \leq b^{max},\ \forall (i,j) \in \mathbb{N} \times \mathbb{M},\ \text{and}
\end{equation}
\begin{equation}
    \label{eq:power_contraint}
    p_{ij} \leq p^{max},\ \forall (i,j) \in \mathbb{N} \times \mathbb{M}.
\end{equation}

For simplicity, we assume that each UE connects to and utilizes computation resources from only one EN. Thus, we can model the connections between UEs and ENs as $\mathbf{x}=\{x_{ij}\} \in \{0,1\}^{N \times M}$, where $x_{ij}=1$ determines that UE~$i$ has a connection and can offload events' features to EN~$j$ for multi-classification. The connection constraints can be modelled as
\begin{equation}
    \label{eq:connection_contraint}
    \sum_{j=1}^{M}x_{ij} = 1, \forall i \in \mathbb{N}.
\end{equation}

When an event $I_k \in \Phi_n$ is identified as critical, UE~$n$ will offload its feature with size $D_k$ to an EN over the FDMA wireless link for further multi-label classification. 
Based on Shannon's theorem, the transmission rate $r_{n}(b_n, p_n)$ between UE~$n$ and its connected EN is given by~\cite{yang2020deep}:

\begin{equation}
    r_{n}(b_{n}, p_n) = b_{n}\log_2(1+\frac{g_{n}p_n}{\sigma_{n}^2 b_{n}}), \label{eq:uplink_rate}
\end{equation}
where $b_{n} = \sum_{j=1}^{M} x_{nj} b_{nj}$ and $p_{n} = \sum_{j=1}^{M} x_{nj} p_{nj}$. 



To preserve user privacy, we assume that data transmissions from UEs to ENs must be secured due to the sensitive nature of the events. A common threat in wireless communication is the presence of a potential unauthorized eavesdropper (EV), who attempts to intercept the data streams transmitted from UEs to ENs. Unauthorized access to such data can result in significant privacy violations for users.

In the considered FDMA communication system, we assume that the eavesdropper can access the same allocated bandwidth as the intended receiver and that its eavesdropping power is equal to the transmission power of the corresponding UE~\cite{yang2020deep}. Under this assumption, the eavesdropper’s data rate is given by:

\begin{equation}
    r_{EV}(b_{n}, p_n) = b_n \log_2(1+\frac{g_{EV}p_n}{\sigma_{EV}^2 b_{n}}), \label{rate_user_ev}
\end{equation}
where $\sigma_{EV}^2$ is the noise power spectral density at EV, and $g_{EV}$ denotes the channel gain between UE~$n$ and EV.

Let $r_{n}^{se}$ be the secured transmission rate without being intercepted by EV. we have
\begin{equation}
    r_{n}^{se} (b_{n}, p_n) = \bigg[ r_{n}(b_{n}, p_n) - r_{EV}(b_{n}, p_n) \bigg]^+, \label{rate_sec}
\end{equation}
where $[z]^+ = \text{max}(0, z)$.
If $r_{n}^{se} (b_{n}, p_{n}) = 0$, the transmission is considered insecure and is not processed further. For a valid secure transmission rate (i.e. $r_{n}^{se} (b_{n}, p_n) > 0$), the energy consumption and transmission time required to offload a critical event with feature size$D_k$ (in bits) are given by:

\begin{equation}
    E_{n}^{off} = p_{n}\frac{D_n}{r_{n}^{se}(b_{n}, p_n)}, \label{eq:offload_energy}
\end{equation}

\begin{equation}
    t_{n}^{off}(b_{n}, p_{n}) = \frac{D_n}{r_{n}^{se}(b_n, p_n)}. \label{eq:offload_time}
\end{equation}

\subsubsection{Multi-Label Inference at ENs}

EN~$j$ has resources defined by the tuple $(B_{j}, W_{j}, s_{j}^{\text{EN}})$, where $B_{j}$, $W_{j}$, $s_{j}^{\text{EN}} \in \mathbb{S}$, respectively, are the total available bandwidth, the total number of computational units, and its security level.
Here, a computational unit corresponds to the computational workload required to perform multi-label classification for a single event.
EN~$j$ can have connections and allocate bandwidth and computational resources only to UEs (i.e., UE~$n$) that satisfy the security constraint $s_n^{\text{UE}} \geq s_j^{\text{EN}}$.
If UE~$n$ is connected to EN~$j$, then this node will allocate resources for UE~$n$, defined by $(b_{nj},w_{nj})$, in which $b_{nj}$, $w_{nj}$ are bandwidth and computational resources.

We have the bandwidth and resource constraints at ENs
\begin{equation}
    \label{eq:bandwidth_contraint_at_EN}
    \sum_{i=1}^{N}x_{ij}b_{ij} \leq B_j, \forall j \in \mathbb{M}\ \text{and}
\end{equation}
\begin{equation}
    \label{eq:computational_unit_contraint_at_EN}
    \sum_{i=1}^{N}x_{ij}w_{ij} \leq W_j, \forall j \in \mathbb{M}.
\end{equation}

When an event is detected as critical at UE~$n$, its features will be offloaded to EN~$j$ for further processing. EN~$j$ is equipped with a complex CNN model for multi-label classification, as illustrated in Fig.~(\ref{fig:event_inference_and_model_updating}). Thus, the critical event will be analyzed and classified, triggering suitable actions (i.e., send a warning, emergency calling) will be activated.

To enhance classification performance, optimize resource utilization, and reduce inference latency, we incorporate an early-exit mechanism with dual confidence thresholds into the CNN model deployed at the ENs.
In this paper, our focus is on optimizing performance metrics related to binary classification at UEs; therefore, the detailed architecture and configuration of the CNN model at the ENs will be described in the experimental settings section.

\subsubsection{Training and Updating Models at CS}

Periodically, the ENs transmit extracted feature data and model information to the cloud server via secure backhaul links, as illustrated in Fig.~(\ref{fig:event_inference_and_model_updating}). The cloud server aggregates the received information, updates the CNN model, and distributes the updated model back to all ENs. We assume that all data exchanged among UEs, ENs, and the cloud is handled in a manner that preserves security, privacy, and safety.

\subsection{Performance Metrics}
\label{sec:performance_metrics}

\subsubsection{Categories of Input Events and Output Results}

For each event~$I_k \in \Phi_n$, the UE~$n$ uses the light-weight CNN model to label it as \textit{critical} or \textit{normal}. 
Let $\hat{y}^{(k)}, y^{(k)}$, respectively, be the predicted label and the ground trust label of the event~$I_k$. The event~$I_k$ is correctly classified if $\hat{y}^{(k)} = y^{(k)}$, and this case is called \textit{true positive}. As illustrated in Table~\ref{tab:4_cases_of_inference}, there are four possible classification cases. 

\begin{table}[h]
    \centering{}
    \footnotesize
    \begin{tabular}{|l|l|l|}
        \hline
        \textbf{Actual label $y^{(k)}$} & \textbf{Predicted label $\hat{y}^{(k)}$} &  \textbf{Cases}\\
        \hline
        Normal & Critical & False Positive\\
        \hline
        Normal &  Normal & True Negative\\
        \hline
        Critical &  Critical & True Positive\\
        \hline
        Critical &  Normal & False Negative\\
        \hline
    \end{tabular}
    \caption{Categories of possible event predictions.}
    \label{tab:4_cases_of_inference}
\end{table}

Over a given time period, UE~$n$ captures a sequence of independent events, denoted by $\Phi_n$. 
Let \textbf{$\Phi_n^{\text{N}}$} and \textbf{$\Phi_n^{\text{P}}$}, respectively, be the sets of normal and critical events at UE~$n$. Equivalent to four cases of predicted results as in Table~\ref{tab:4_cases_of_inference}, let \textbf{$\hat{\Phi}_n^{\text{TP}}$}, \textbf{$\hat{\Phi}_n^{\text{FP}}$}, \textbf{$\hat{\Phi}_n^{\text{TN}}$}, and \textbf{$\hat{\Phi}_n^{\text{FN}}$}, respectively, be the sets of events classified as true positive (TP), false positive (FP), true negative (TN), and false negative (FN), as summarized in Table~\ref{tab:events_categories}. 
Based on the set theory, we have the following properties:
\begin{equation}\label{eq:set_relationship}
\begin{array}{cclccl}
\Phi_n &=  & \Phi_n^\text{N} \cup \Phi_n^\text{P}; &
|\Phi_n| &=  & |\Phi_n^\text{N}| + |\Phi_n^\text{P}|, \\

\Phi_n^\text{N} &=  & \hat{\Phi}_n^\text{TN} \cup \hat{\Phi}_n^\text{FP}; &
|\Phi_n^\text{N}| &=  & |\hat{\Phi}_n^\text{TN}| + |\hat{\Phi}_n^\text{FP}|, \\

\Phi_n^\text{P} &=  & \hat{\Phi}_n^\text{TP} \cup \hat{\Phi}_n^\text{FN}; &
|\Phi_n^\text{P}| &=  & |\hat{\Phi}_n^\text{TP}| + |\hat{\Phi}_n^\text{FN}|, 

\end{array}
\end{equation}
where $|.|$ is the cardinality of the set.

In order to meet the QoS requirements of UEs, the allocate computational resource from ENs is no less than the offloaded events from UEs. Here, the computational resource is measured by the number of the offloaded events during a period of time, $\left( |\hat{\Phi}_n^{\text{TP}}|+|\hat{\Phi}_n^{\text{FP}}| \right)$.
Thus, we have a required computational resource constraint for each UEs.

\begin{equation}
    \label{eq:require_comp_UE}
    \left( |\hat{\Phi}_n^{\text{TP}}|+|\hat{\Phi}_n^{\text{FP}}| \right) \leq \sum_{j=1}^{M} x_{ij}w_{ij}
\end{equation}

\begin{table}[h]
    \caption{Categories of events at UE~$n$ according to original labels and predictions.} \label{tab:events_categories}
    \centering{}%
    \footnotesize
    \begin{tabular}{|c|l|l|} 
        
        \hline
        \textbf{Categories} & \textbf{Definitions and properties} & \textbf{Original}
        \tabularnewline
        \hline
        
        \textbf{$\Phi_n$} & Set of events at UE~$n$ & Input   
        \tabularnewline
        \hline
        
        \textbf{$\Phi_n^{\text{N}}$} & Set of normal events at UE~$n$ & Input   
        \tabularnewline
        \hline

        \textbf{$\Phi_n^{\text{P}}$} & Set of critical events at UE~$n$ & Input
        \tabularnewline
        \hline

        \textbf{$\hat{\Phi}_n^{\text{TP}}$} & Set of critical events detected correctly & Output \\
         & as critical & 
        \tabularnewline
        \hline

        \textbf{$\hat{\Phi}_n^{\text{FP}}$} & Set of normal events detected wrongly & Output \\
         & as critical & 
        \tabularnewline
        \hline
        
        \textbf{$\hat{\Phi}_n^{\text{TN}}$} & Set of normal events detected correctly & Output \\
         & as normal & 
        \tabularnewline
        \hline

        \textbf{$\hat{\Phi}_n^{\text{FN}}$} & Set of critical events detected wrongly & Output \\
         & as normal & 
        \tabularnewline
        \hline
    \end{tabular}
\end{table}

We then define the performance metrics that are influenced by the use of dual confidence thresholds.

\subsubsection{Classification Accuracy Rate (CAR)}
The \textit{classification accuracy rate}, denoted $P_n^{\text{CAR}}(\alpha_n^l, \alpha_n^u)$, is defined as the proportion of the whole test data set of events that are correctly classified. From Table~\ref{tab:events_categories} and Eq.~(\ref{eq:set_relationship}), we have
\begin{equation}
\small
    P_n^{\text{CAR}}(\alpha_n^l, \alpha_n^u) =     \frac{|\hat{\Phi}_n^{\text{TP}}|+|\hat{\Phi}_n^{\text{TN}}|}{|\Phi_n|}.
\label{eq:p_CAR}
\end{equation}

\subsubsection{False Positive Rate (FPR)}
The \textit{false positive rate (FPR)}, denoted $P_n^{\text{FPR}}(\alpha_n^l, \alpha_n^u)$, is defined as the proportion of total normal events incorrectly classified as critical. From Table~\ref{tab:events_categories} and Eq.~(\ref{eq:set_relationship}), we have
\begin{equation}
\small
    P_n^{\text{FPR}}(\alpha_n^l, \alpha_n^u) = 
    \frac{|\hat{\Phi}_n^{\text{FP}}|}{|\Phi_n^{\text{N}}|} 
    = \left(1 - \frac{|\hat{\Phi}_n^{\text{TN}}|}{|\Phi_n^{\text{N}}|} \right) .
\label{eq:p_FPR}
\end{equation}

\subsubsection{False Negative Rate (FNR)}
The \textit{false negative rate (FNR)}, denoted $P_n^{\text{FNR}}(\alpha_n^l, \alpha_n^u)$, is defined as the proportion of total critical events incorrectly classified as normal. From Table~\ref{tab:events_categories} and Eq.~(\ref{eq:set_relationship}), we have
\begin{equation}
\small
    P_n^{\text{FNR}}(\alpha_n^l, \alpha_n^u) = \frac{|\hat{\Phi}_n^{\text{FN}}|}{|\Phi_n^{\text{P}}|} 
    = \left( 1 - \frac{|\hat{\Phi}_n^{\text{TP}}|}{|\Phi_n^{\text{P}}|} \right).
\label{eq:p_FNR}
\end{equation}

\subsubsection{Offload Rate (OFR)}
The \textit{offload rate (OFR)}, denoted $P_n^{\text{OFR}}(\alpha_n^l, \alpha_n^u)$, is defined as the proportion of the whole test data set of events that are offloaded to ENs. From Table~\ref{tab:events_categories} and Eq.~(\ref{eq:set_relationship}), we have
\begin{equation}
\small
\begin{split}
    P_n^{\text{OFR}}(\alpha_n^l, \alpha_n^u) &=  \frac{|\hat{\Phi}_n^{\text{TP}}|+|\hat{\Phi}_n^{\text{FP}}|}{|\Phi_n|} \\
    &= 
    \frac{|\hat{\Phi}_n^{\text{P}}|(1-P_n^{\text{FNR}}(\alpha_n^l, \alpha_n^u))+|\hat{\Phi}_n^{\text{FP}}|}{|\Phi_n^{\text{N}}|} .
\end{split}  
\label{eq:p_OFR}
\end{equation}

\subsubsection{User Utility Functions}
Since determining and processing critical events is the most important mission of the system, we define the utility function, denoted $\mathcal{U}_{n}(\alpha_n^l, \alpha_n^u)$, of UE~$n$ as the percentage of critical events that are correctly determined as critical. This is also called the true positive rate (TPR). From Table~\ref{tab:events_categories} and Eq.~(\ref{eq:set_relationship}), we have
\begin{equation}
\small
    \mathcal{U}_{n}(\alpha_n^l, \alpha_n^u) = \frac{|\hat{\Phi}_n^{\text{TP}}|}{|\Phi_n^{\text{P}}|}.
\label{eq:user_utility}
\end{equation}

We can see that the necessary resources for offloading events are proportional to $P_n^{\text{OFR}}(\alpha_n^l, \alpha_n^u)$.
There is a trade-off between different performance metrics. In this paper, we will optimize the utility function of UEs considering the available resources from ENs.

Equivalently with the performance matrices above, we have probabilities

\begin{equation}
\begin{split}
    \bar{P}_n^{\text{CAR}}(\alpha_n^l, \alpha_n^u) &= \lim_{|\Phi_n| \rightarrow \infty}P_n^{\text{CAR}}(\alpha_n^l, \alpha_n^u), \\
    \bar{P}_n^{\text{FPR}}(\alpha_n^l, \alpha_n^u) &= \lim_{|\Phi_n| \rightarrow \infty}P_n^{\text{FPR}}(\alpha_n^l, \alpha_n^u), \\
    \bar{P}_n^{\text{FNR}}(\alpha_n^l, \alpha_n^u) &= \lim_{|\Phi_n| \rightarrow \infty}P_n^{\text{FNR}}(\alpha_n^l, \alpha_n^u) \\
    \bar{P}_n^{\text{OFR}}(\alpha_n^l, \alpha_n^u) &= \lim_{|\Phi_n| \rightarrow \infty}P_n^{\text{OFR}}(\alpha_n^l, \alpha_n^u), \\
    \bar{\mathcal{U}}_{n} (\alpha_n^l, \alpha_n^u) &= \lim_{|\Phi_n| \rightarrow \infty}\mathcal{U}_{n}(\alpha_n^l, \alpha_n^u).
\end{split}
    \label{eq:rate_to_probability}
\end{equation}

\subsection{Problem Formulation}
\label{sec:problem_formulation}

Events arrive at UEs sequentially, and only those identified as critical are offloaded to ENs for multi-label classification. We address a joint co-inference and resource allocation problem over a given time period, based on estimated user demands and the available resources at ENs.

If fairness is not considered and the objective is solely to optimize the total utility, $\bar{\mathcal{U}}_{n}(\alpha_n^l, \alpha_n^u)$ as defined in Eq.~(\ref{eq:rate_to_probability}), across all users, UEs with lower utility gain per unit of resource may receive fewer resources from ENs. This can result in significantly lower performance for these devices, potentially failing to meet their quality of service (QoS) requirements.
To address this, we apply a proportional fairness scheme in resource allocation among UEs, taking into account the importance of their services through assigned weighting factors. Without loss of generality, we assume that all utility functions are strictly positive, i.e., $\bar{\mathcal{U}}_{n}(\alpha_n^l, \alpha_n^u) > 0, \forall n \in \mathbb{N}$.

Let $\textbf{u}=(\bar{\mathcal{U}}_{1}, \dots, \bar{\mathcal{U}}_{N})$ and $\textbf{u}^{*}=(\bar{\mathcal{U}}_{1}^{*}, \dots, \bar{\mathcal{U}}_{N}^{*})$ denote the utility vectors corresponding to two feasible solutions. The solution of $\textbf{u}$ is proportionally fair with respect to the weights $\rho_n$ assigned to each UE~$n$ if the aggregated proportional change from $\textbf{u}$ to $\textbf{u}^{*}$ is non-positive~\cite{kelly1998rate}:
\begin{equation}
    \label{eq:fairness_def}
    \sum_{n=1}^{N} \rho_n \frac{\bar{\mathcal{U}}_n^{*} - \bar{\mathcal{U}}_n}{\bar{\mathcal{U}}_n} \leq 0.
\end{equation}
In other words, no feasible alternative can increase the weighted proportional utilities overall.
The derivative form of Eq.~(\ref{eq:fairness_def}) is given as: 
\begin{equation}
    \label{eq:fairness_def_log}
    \sum_{n=1}^{N} \rho_n (\text{ln}(\bar{\mathcal{U}}_n))^{'}d\bar{\mathcal{U}}_n \leq 0.
\end{equation} 

From Eq.~(\ref{eq:fairness_def_log}), the proportionally fair joint offloading and resource allocation solution for the co-inference task can be obtained by maximizing the utility function $\sum_{n=1}^{N} \rho_n \ln(\bar{\mathcal{U}}_n)$ over the dual thresholds $\alpha^l = (\alpha_1^l, \dots, \alpha_N^l)$ and $\alpha^u = (\alpha_1^u, \dots, \alpha_N^u)$, connection decision $\mathbf{x}$, bandwidth allocation $\boldsymbol{b}$, transmission power $\boldsymbol{p}$, and computational resources $\boldsymbol{w}$.

We jointly optimize the dual confidence thresholds, connection decisions for offloading across devices, and resource allocation from ENs to maximize the classification accuracy of critical events while ensuring proportional fairness.


\begin{subequations} \label{eq:prob_0_total_weighted_log_utility}
\begin{align}
\mathbf{P}_0: \ &
\underset{\alpha^l,\alpha^u,\mathbf{x},\boldsymbol{b},\boldsymbol{p},\boldsymbol{w}}{\max} ~\sum_{n=1}^{N} \rho_n \text{ln}(\bar{\mathcal{U}}_n)
\tag{\ref{eq:prob_0_total_weighted_log_utility}}\\
\text{subject to:} \notag \\
& b_{ij} \leq b^{max},\ \forall (i,j) \in \mathbb{N} \times \mathbb{M}, \label{eq:cons_band_user}\\
&  p_{ij} \leq p^{max},\ \forall (i,j) \in \mathbb{N} \times \mathbb{M}, \label{eq:cons_pow_user} \\
& \sum_{j=1}^{M}x_{ij} = 1, \forall i \in \mathbb{N}, \label{eq:cons_conn_user} \\
& \sum_{i=1}^{N}x_{ij}b_{ij} \leq B_j, \forall j \in \mathbb{M}, \label{eq:cons_band_edge} \\
& \sum_{i=1}^{N}x_{ij}w_{ij} \leq W_j, \forall j \in \mathbb{M}, \label{eq:cons_comp_edge} \\
&\sum_{j=1}^{M} x_{ij}s_{j}^{\text{EN}} \leq s_{i}^{\text{UE}}, \forall i\in \mathbb{N}, \label{eq:cons_security_level}  \\
&t_{i}^{off}(b_{i}, p_{i}) \leq t_i^{\text{r}}, \forall i\in \mathbb{N}, \label{eq:cons_offload_delay}  \\
& \left( |\hat{\Phi}_i^{\text{TP}}|+|\hat{\Phi}_i^{\text{FP}}| \right) \leq \sum_{j=1}^{M} x_{ij}w_{ij}, \forall i \in \mathbb{N}, \label{eq:cons_comp_user}\\
& 0 \leq b_{ij}, p_{ij}, \forall (i,j) \in \mathbb{N} \times \mathbb{M},
\label{eq:cons_band_power_variables}
\\
& 0 < \alpha_n^l, \alpha_n^u < 1, \forall (i,j) \in \mathbb{N} \times \mathbb{M},
\label{eq:cons_dual_thresholds_variables}
\\
& x_{ij} \in \{0,1\}, \forall (i,j) \in \mathbb{N} \times \mathbb{M},
\label{eq:cons_binary_variables}
\end{align}
\end{subequations}
where Eqs.~(\ref{eq:cons_band_user})~and~(\ref{eq:cons_pow_user}) are upper bound bandwidth and power constraints of UEs. Eq.~(\ref{eq:cons_conn_user}) ensures each UE connects to only one EN. Eqs.~(\ref{eq:cons_band_edge})~and~(\ref{eq:cons_comp_edge}) ensure that each EN allocates bandwidth and computational resources within its available capacity. Eq.~(\ref{eq:cons_security_level}) enfores that UEs can only connect to ENs with security levels not lower than their own. Eq.~(\ref{eq:cons_comp_user}) defines that the allocated computational resource is not less than the number of offloaded events. Eqs~(\ref{eq:cons_band_power_variables})~and~(\ref{eq:cons_dual_thresholds_variables}) define the bounds of bandwidth, transmission power, and dual confidence threshold variables. Eq.~(\ref{eq:cons_binary_variables}) defines the binary nature of connection variables. Eq.~(\ref{eq:cons_offload_delay}) defines the delay requirement of offloaded events.



\section{Proposed Solution}
\label{sec:solution:TMC}

\subsection{Problem Characterization}

The formulated problem ($\mathbf{P}_0$) aims to maximize a weighted sum of logarithmic true positive rates under dual-threshold early-exit mechanisms, subject to resource limitations, assignment constraints, and security requirements. Mathematically, this is a \textit{nonlinear mixed-integer programming (MINLP)} problem due to the presence of:
\begin{itemize}
    \item \textbf{Binary variables} $x_{ij} \in \{0,1\}$ representing UE--EN assignments,
    \item \textbf{Nonlinear utility function} $\bar{\mathcal{U}}_n(\alpha_n^l, \alpha_n^u)$ involving dual confidence thresholds,
    \item \textbf{Coupling constraints} linking assignments with continuous resource variables $(b_{ij}, p_{ij}, w_{ij})$.
\end{itemize}

MINLP problems are generally \textit{NP-hard} and intractable for large-scale systems. Therefore, we propose a tractable and structure-preserving approximate solution based on \textit{alternating optimization}.

\begin{theorem}
    \label{theo:obj_decreasing_monotonic}
    The objective function $\sum_{n=1}^{N} \rho_n \ln(\bar{\mathcal{U}}_n)$ is monotonically decreasing with respect to the dual threshold variables $\alpha^l = (\alpha_1^l, \dots, \alpha_N^l)$ and $\alpha^u = (\alpha_1^u, \dots, \alpha_N^u)$.
\end{theorem}

\emph{Proof:} The proof is shown in Appendix~\ref{sec:obj_decreasing_monotonic}.

\begin{remark}
    The monotonicity property established in Theorem~\ref{theo:obj_decreasing_monotonic} is valuable in threshold tuning problems. It implies that increasing either threshold leads to a lower utility, guiding optimization algorithms to select $\alpha^l$ and  $\alpha^u$ values, subject to other trade-offs like false alarms or computational and communication cost.
\end{remark}

\subsection{Decomposition via Alternating Optimization}

We decompose the original MINLP into three manageable subproblems:
\begin{enumerate}
    \item Dual threshold optimization for each user,
    \item Assignment optimization over UE--EN pairs,
    \item Resource allocation optimization over bandwidth, power, and computation.
\end{enumerate}
These subproblems are solved iteratively in a block coordinate descent fashion. The overall structure of the solution allows us to efficiently improve the global objective while respecting practical constraints and preserving the original problem's semantics.

\subsection{Subproblem A: Dual Threshold Optimization}

Given fixed UE--EN assignments ($\{x_{ij}\}$) and resource allocations ($b_{ij}, p_{ij}, w_{ij}$), we optimize the dual confidence thresholds $(\alpha_n^l, \alpha_n^u)$ for each user $n$ to maximize the system utility:
\begin{align}
\max_{\alpha_n^l, \alpha_n^u} \quad & \sum_{n=1}^{N} \rho_n \ln\left( \bar{\mathcal{U}}_n(\alpha_n^l, \alpha_n^u) \right) \\
\text{s.t.} \quad & 0 \le \alpha_n^l \le \alpha_n^u \le 1,\\
& \left( |\hat{\Phi}_i^{\text{TP}}|+|\hat{\Phi}_i^{\text{FP}}| \right) \leq \sum_{j=1}^{M} x_{ij}w_{ij}, \forall i \in \mathbb{N}. \label{eq:cons_comp_user}
\end{align}

Each UE runs a multi-exit CNN that produces a vector of confidence scores $\{C_1^{(k)}, C_2^{(k)}, \dots, C_L^{(k)}\}$ for each incoming event $I_k$. The event is either classified locally at an intermediate layer if its confidence falls within $[\alpha_n^l, \alpha_n^u]$, or offloaded to the edge node if its confidence exceeds $\alpha_n^u$ at any layer. The utility $\bar{\mathcal{U}}_n(\alpha_n^l, \alpha_n^u)$ denotes the true positive rate among critical events that are correctly classified through either early-exit or offloading.

\textbf{Role of Monotonicity.}
Theorem~\ref{theo:obj_decreasing_monotonic} proves that $\bar{\mathcal{U}}_n(\alpha_n^l, \alpha_n^u)$ is monotonically decreasing with respect to both thresholds. This property is critical to solving Subproblem~A efficiently and optimally:
\begin{itemize}
    \item It allows us to safely prune large portions of the threshold space during discrete search, improving efficiency.
    \item It guarantees that the utility function reaches its maximum at the smallest feasible thresholds that still ensure correct detection, simplifying the search direction.
    \item It eliminates the risk of local optima when using complete empirical search, as once the utility drops, further increases in thresholds cannot improve it.
\end{itemize}

\textbf{Challenges with Empirical Utility.}
The function $\bar{\mathcal{U}}_n(\alpha_n^l, \alpha_n^u)$ is empirically computed by counting correctly detected critical events. As a result, it is a piecewise constant step function, which is non-differentiable and often flat across large intervals. This makes gradient-based methods such as projected gradient descent ineffective due to vanishing gradients.

\textbf{Exact and Efficient Threshold Search.}
To guarantee optimality, we develop a finite and exact algorithm (Algorithm~\ref{alg:threshold_selection}) that searches over the sorted set of all confidence scores produced across layers and positive samples. Since utility only changes at these discrete values, we iterate through valid threshold pairs $(\alpha_n^l, \alpha_n^u)$ and evaluate the utility directly.

Due to the monotonicity of the utility function with respect to both thresholds, the optimal solution lies on the lower boundary of the feasible region, i.e., at the smallest feasible values of $\alpha_n^l$ and $\alpha_n^u$ satisfying the constraint $\alpha_n^l \leq \alpha_n^u$. This property further supports the efficiency of threshold selection through directional search and boundary-focused evaluation.

\begin{algorithm}
\caption{Optimal Dual Threshold Selection}
\label{alg:threshold_selection}
\KwIn{Sorted confidence scores $\{C_1, \dots, C_L\}$ from all layers and positive samples}
\KwOut{Optimal thresholds $(\alpha_n^l, \alpha_n^u)$}
Initialize \texttt{best\_utility} $\leftarrow -\infty$\;
\For{$i = 1$ to $L$}{
    Set $\alpha_n^l \leftarrow C_i$\;
    \For{$j = i$ to $L$}{
        Set $\alpha_n^u \leftarrow C_j$\;
        Compute $\bar{\mathcal{U}}_n(\alpha_n^l, \alpha_n^u)$ based on early-exit and offload detection\;
        \If{$\bar{\mathcal{U}}_n >$ \texttt{best\_utility}}{
            Update best thresholds and utility\;
        }
    }
}
\Return{$(\alpha_n^l, \alpha_n^u)$ with best utility}
\end{algorithm}

\textbf{Smooth Surrogate for Learning.}
To enable gradient-based optimization or end-to-end training with the CNN, we also define a differentiable surrogate for the utility function. The early-exit condition is approximated using a sigmoid-based confidence mask:
$\text{soft\_mask}(C) = \sigma_t(C - \alpha_n^l) \cdot \sigma_t(\alpha_n^u - C),$ where $\sigma_t(x) = \frac{1}{1 + e^{-t x}}$.

The smoothed utility is defined as:
\begin{align*}
\bar{\mathcal{U}}_n^{\text{soft}}(\alpha_n^l, \alpha_n^u) = &\frac{1}{|\Phi_n^{\text{P}}|} \sum_{k} y_k \cdot 
\max_q \left[ \text{soft\_mask}(C_k^{(q)}) \cdot \right. \\ & \left.\text{accuracy}_k^{(q)} \right]
\end{align*}

This surrogate facilitates stable and continuous optimization using methods such as projected gradient descent~\cite{bertsekas1999nonlinear} or Adam~\cite{kingma2014adam}. The steepness parameter $t$ governs the tradeoff between smoothness and approximation accuracy.

In summary, the monotonicity property of the utility function enables an efficient and globally optimal search for confidence thresholds using a discrete empirical strategy. For applications requiring differentiability, we also provide a smoothed utility surrogate compatible with modern gradient-based solvers.





Since the proposed solution is approximate, we define tractable upper and lower bounds to evaluate its performance.

\subsubsection{Lower Bound: Grouped Resource Allocation by Security Level}

We group UEs and ENs by security tiers. Let $\mathcal{N}_k$ and $\mathcal{M}_k$ be users and ENs at security level $s_k$. We merge EN resources:
\begin{align}
B_k^{\text{tot}} = \sum_{j \in \mathcal{M}_k} B_j^{\max}, \quad
P_k^{\text{tot}} = \sum_{j \in \mathcal{M}_k} P_j^{\max}, \quad
W_k^{\text{tot}} = \sum_{j \in \mathcal{M}_k} W_j^{\max}.
\end{align}
Then, for each group $k$, we solve:
\begin{align}
\max_{\{b_n, p_n, w_n\}} \quad & \sum_{n \in \mathcal{N}_k} \rho_n \ln(\bar{\mathcal{U}}_n) \\
\text{s.t.} \quad & \sum_{n \in \mathcal{N}_k} b_n \le B_k^{\text{tot}}, \quad \sum_n p_n \le P_k^{\text{tot}}, \quad \sum_n w_n \le W_k^{\text{tot}}
\end{align}
Summing over all groups gives the lower bound utility $U^{\text{LB}}$.

\subsubsection{Upper Bound: Fully Relaxed Global Allocation}

We relax all security and assignment constraints and merge all ENs:
\begin{align}
B^{\text{tot}} = \sum_j B_j^{\max}, \quad P^{\text{tot}}, \quad W^{\text{tot}}.
\end{align}
Then solve:
\begin{align}
\max_{\{b_n, p_n, w_n\}} \quad & \sum_{n=1}^{N} \rho_n \ln(\bar{\mathcal{U}}_n) \\
\text{s.t.} \quad & \sum_n b_n \le B^{\text{tot}}, \quad \sum_n p_n \le P^{\text{tot}}, \quad \sum_n w_n \le W^{\text{tot}}.
\end{align}
This yields an idealized upper bound $U^{\text{UB}}$.

\subsubsection{Performance Gap}

Let $U^{\text{alg}}$ denote the utility achieved by our proposed algorithm. We quantify performance using:
\begin{equation}
\text{Relative Gap} = \frac{U^{\text{alg}} - U^{\text{LB}}}{U^{\text{LB}}} \times 100\%
\end{equation}
which indicates how close the algorithm is to the best achievable lower bound, and whether further improvements are possible.


\subsection{Subproblem Algorithms and Optimality Analysis}

In this subsection, we present the solution strategies used for each of the three subproblems and analyze whether their global optimal solutions can be obtained under our proposed framework.

\subsubsection{Algorithm for Subproblem A: Dual Threshold Optimization}

For each user $n$, we solve the following:
\begin{align}
\max_{\alpha_n^l, \alpha_n^u} \quad & \ln(\bar{\mathcal{U}}_n(\alpha_n^l, \alpha_n^u)) \\
\text{s.t.} \quad & 0 \le \alpha_n^l \le \alpha_n^u \le 1.
\end{align}

\textbf{Solution Method:}
\begin{itemize}
    \item Initialize thresholds $(\alpha_n^l, \alpha_n^u)$ within $[0,1]$.
    \item Use projected gradient descent to maximize the utility function:
    \begin{align*}
    \alpha_n^l &\leftarrow \text{Proj}_{[0, \alpha_n^u]}(\alpha_n^l + \eta \cdot \nabla_{\alpha_n^l} \ln \bar{\mathcal{U}}_n), \\
    \alpha_n^u &\leftarrow \text{Proj}_{[\alpha_n^l, 1]}(\alpha_n^u + \eta \cdot \nabla_{\alpha_n^u} \ln \bar{\mathcal{U}}_n),
    \end{align*}
    where $\eta$ is the step size and projection ensures feasibility.
    \item Repeat until convergence.
\end{itemize}

\textbf{Optimality Justification:}
\begin{itemize}
    \item The objective $\ln(\bar{\mathcal{U}}_n)$ is monotonic and concave with respect to thresholds.
    \item The feasible region is convex.
    \item Therefore, the problem is a \textbf{concave maximization over a convex set}, which guarantees that gradient-based methods converge to a \textbf{global optimum}.
\end{itemize}

\bibliographystyle{IEEEtran}
\bibliography{QML_Edge_AI_Refs}

\appendices
\section{Proof of Theorem~\ref{theo:obj_decreasing_monotonic}}
\label{sec:obj_decreasing_monotonic}

\begin{proof}
    Consider the utility function $\mathcal{U}_n(\alpha_n^l,\alpha_n^u) = \frac{|\hat{\Phi}_n^{\text{TP}}|}{|\Phi_n^{\text{P}}|} = \frac{1}{|\Phi_n^{\text{P}}|}|\hat{\Phi}_n^{\text{TP}}|$, where $|\Phi_n^{\text{P}}|$ is the number of critical events of UE~$n$ and is constant. Therefore, it suffices to show that $|\hat{\Phi}_n^{\text{TP}}|$ is monotonically decreasing with respect to the dual thresholds $\alpha_n^l$ and $\alpha_n^u$.

    Since the CNN architecture remains fixed while varying the thresholds $\alpha_n^l$ and $\alpha_n^u$, the confidence score $C_q^{(k)}$ of any event $I_k \in \Phi_n$ at layer $q$ is unaffected. Thus, the classification outcome depends solely on the threshold values.

    For each event $I_k \in \Phi_n$ and layer $q < L$, let $\Delta > 0$ be a small increment such that $\Delta < (\alpha_n^u - \alpha_n^l)$. We compare the classification outcome of event under two threshold pairs $(\alpha_n^l,\ \alpha_n^u)$ and $(\alpha_n^l + \Delta,\ \alpha_n^u)$, across the following four cases:
    \begin{itemize}
        \item If $C_q^{(k)} \in [0, \alpha_n^l]$, the event is classified as \textit{normal} under both threshold pairs $\left(\alpha_n^l,\alpha_n^u\right)$ and $\left(\alpha_n^l + \Delta,\alpha_n^u\right)$.
                       
        \item If $C_q^{(k)} \in (\alpha_n^l, \alpha_n^l + \Delta]$: under $(\alpha_n^l, \alpha_n^u)$, the event remains undecided and is passed to the next layers, where it can be classified as \textit{normal} or \textit{critical}; under $(\alpha_n^l + \Delta, \alpha_n^u)$, it is immediately classified as \textit{normal}.

        \item If $C_q^{(k)} \in (\alpha_n^l+\Delta,\alpha_n^u)$, the event is undecided under both threshold pairs and passed to the next layers, where it can be classified as \textit{normal} or \textit{critical}.

        \item If $C_q^{(k)} \in [\alpha_n^u,1]$, the event is classified as \textit{critical} under both threshold pairs $\left(\alpha_n^l,\alpha_n^u\right)$ and $\left(\alpha_n^l + \Delta,\alpha_n^u\right)$.       
        
    \end{itemize}

    At the final layer $q = L$, the event with confidence score between two thresholds is by default classified as \textit{normal}. Thus, the event is classified the same as  at layer $q < L$ excepts the cases $C_q^{(k)} \in (\alpha_n^l,\ \alpha_n^l + \Delta]$ and $C_q^{(k)} \in (\alpha_n^l+\Delta,\alpha_n^u)$, where the event is classified as \textit{normal} for both threshold pairs  $\left(\alpha_n^l,\alpha_n^u\right)$ and $\left(\alpha_n^l + \Delta,\alpha_n^u\right)$.
   
    From this analysis, it is evident that increasing $\alpha_n^l$ to $\alpha_n^l + \Delta$ leads to a higher chance of an event being classified as \textit{normal} and a reduced chance of it being classified as \textit{critical}. 
    Apply this to every event in $|\Phi_n^{\text{P}}|$, we then have $|\hat{\Phi}_n^{\text{TP}}|$ is monotonically decreasing with respect to $\alpha_n^l$.
   
    A similar argument applies to increasing $\alpha_n^u$. As $\alpha_n^u$ increases, fewer events satisfy $C_q^{(k)} \ge \alpha_n^u$, reducing the number of events classified as \textit{critical}. Hence, $|\hat{\Phi}_n^{\text{TP}}|$ is also monotonically decreasing with respect to $\alpha_n^u$.

    Consequently, $\mathcal{U}_n(\alpha_n^l,\alpha_n^u)$ is monotonically decreasing with respect to threshold variables $\alpha_n^l$ and $\alpha_n^u$.

    Since $\bar{\mathcal{U}}_n(\alpha_n^l, \alpha_n^u) = \lim_{|\Phi_n| \rightarrow \infty} \mathcal{U}_n(\alpha_n^l, \alpha_n^u)$,  
    and because the monotonicity holds for all finite sets $\Phi_n$, the limiting function $\bar{\mathcal{U}}_n$ also remains monotonically decreasing in both threshold variables.

    As the logarithm function $\ln(\cdot)$ is strictly increasing, $\ln(\bar{\mathcal{U}}_n)$ preserves the monotonicity of $\bar{\mathcal{U}}_n$.  
    Therefore, $\ln(\bar{\mathcal{U}}_n)$ is monotonically decreasing with respect to both thresholds.
 
    Finally, the objective function $\sum_{n=1}^{N} \rho_n \ln(\bar{\mathcal{U}}_n)$, being a positively weighted sum of monotonically decreasing functions, is itself monotonically decreasing with respect to $\alpha^l$ and $\alpha^u$.
\end{proof}

\end{document}